\documentclass[final,11p,times,authoryear]{article}

\usepackage{soul}
\usepackage{amsfonts}
\usepackage{amsmath}
\allowdisplaybreaks[4]
\usepackage{amssymb}
\usepackage{euscript}
\usepackage{color}
\usepackage{tensor}
\usepackage{mathrsfs}
\usepackage{float}
\usepackage{graphicx}
\usepackage{caption}
\usepackage{subcaption}

\usepackage{hyperref}

\usepackage[style = numeric]{biblatex}
\usepackage{parskip}
\usepackage{geometry}
\usepackage[toc,page]{appendix}
\usepackage{authblk}

\begin{document}
\font\myfont=cmr12 at 20pt



\title{{\Huge{\bf{The Discovery of Neptune Revisited}}}}


\author[1]{Gabriel Rodríguez-Moris}
\author[2, 3, 4]{José A. Docobo}

\font\myfont=cmr12 at 10pt

\affil[1]{\myfont Lohrmann-Observatorium, Technische Universität Dresden (TUD), August-Bebel-Straße 30, Dresden, 01219, Germany}
            
\affil[2]{\myfont Centro de Investigación e Tecnoloxía Matemática de Galicia (CITMAga), Santiago de Compostela, 15782, Galiza, Spain}

\affil[3]{\myfont Observatorio Astronómico Ramón María Aller, Universidade de Santiago de Compostela (USC) Avenida das Ciencias s/n, Campus Vida, Santiago de Compostela, 15782, Galiza, Spain}

\affil[4]{\myfont Real Academia de Ciencias de Zaragoza, Facultad de Ciencias, C/ Pedro Cerbuna 12, 50009, Zaragoza, Spain}

\maketitle
\begin{abstract}
The study of the differences detected between the observed and the predicted positions of Uranus taking only the ancient planets into account led to the discovery of planet Neptune in 1846. This event remains one of the best accomplishments ever achieved in the history of Astronomy and Classical Mechanics. In this paper, we study the perturbations in the orbit of Uranus due to Neptune and its effects from a modern numerical point of view of the $N$-body problem. The effects induced by Pluto in the orbit of Neptune, as the historical search for a ninth planet in the Solar System (recently boostered again with the hypothesis of the so-called Planet Nine) back in the days was propelled by some supposed small inconsistencies in the orbit of the ice giants, are also analyzed.

\end{abstract}










\section{Introductory and historical remarks}
\label{Intro}

\noindent
The night sky has captivated the curiosity of human beings since ancient times, which has led Astronomy to be regarded as the first scientific field to ever be investigated. Already very early civilizations noticed that, besides the Sun and the Moon, five of the "stars" visible to the naked eye appeared to noticeably vary their position on the sky with respect to the other stars, after a relatively short period of time. Due to this erratic motion, these "stars" were coined "planets". Later, in the modern era, it was well-established that, like the Earth, they are also orbiting our host star, the Sun, and that they are in fact much closer to us than the other "background" stars. These are the classical "ancient" planets: Mercury, Venus, Mars, Jupiter and Saturn, in increasing order of distance to the Sun.

It was not until 1781 that, using a telescope, William Herschel discovered Uranus by chance, a planet beyond Saturn whose apparent magnitude made it appear extremely faint to the naked eye and only in very favourable conditions, even though afterwards it was noticed that it had already been observed in telescopic observations some times before in history, but was confused with a different kind of astronomical body. Following the discovery of Uranus, astronomers tried to predict its future positions using perturbation theory applied to the exactly integrable Newtonian $2$-body problem in Celestial Mechanics. However, these were rapidly found to differ from the actual observed positions. These mismatches led some astronomers of the time, such as the director of the Paris Observatory Alexis Bouvard, to propose, among other hypotheses, the existence of a body farther away than Uranus whose perturbations on its orbit were responsible for them. Within this framework, John Couch Adams and Urbain Le Verrier reached, independently but using essentially the same perturbation theory techniques, a predicted position on the celestial sphere for the supposed perturber\footnote{In order to calculate its distance from the Sun, they made use of the Titius-Bode law, which back in the days was a good estimator even though nowadays we know it is not correct.}, which was found to be very close to the actual position spotted in September of 1846 by Johann Gottfried Galle and Heinrich Louis d'Arrest at the Berlin Observatory, a truly remarkable achievement for the epoch. This led to the discovery of planet Neptune, the eighth planet of the Solar System. As it happened with Uranus, we currently know that Neptune had also been observed before its recoginition as a planet, \textit{e.g.} by Galileo at the end of 1608 \cite{davor1, Krajnovi__2016, Smart}. The discovery quickly aroused the interest of astronomers of all over the world, and subsequents observations and studies were correspondingly reported in renowned astronomical journals such as \textit{Astronomische Nachrichten}; see for instance the note by the Astronomer Royal George Airy \cite{AN}.

Later, different mistmatches regarding the perihelion precession of Mercury led Le Verrier to additionally propose an inner planet to it; however, the source of this problem was a more fundamental one: it required the formalism of General Relativity, as a consequence of the stronger gravitational field in regions close to the Sun. This was the first astronomical problem which seriously required the relativistic formulation of the gravitational interaction. In our days, it is quite noteworthy that many subtle relativistic effects and their corresponding physical interpretations (see \textit{e.g.} \cite{AGR} for detailed developments in the field of applied General Relativity) are needed to explain and model observational data from high-precision missions like Gaia \cite{Gaia}, which is currently providing the greatest astrometric catalog (of $\mathcal{O}(10^9)$ sources) ever built, with striking accuracies of microarcseconds and potentially improving in future potential missions like GaiaNIR \cite{GaiaNIR}, allowing an unprecedented understanding of many topics in several branches of Astronomy and Astrophysics which is only increasing, and this only a century and a half after the discovery of the eighth planet of the Solar System! Indeed, the $5$ typical astrometric parameters (namely parallax, angular positions in the celestial sphere and the corresponding proper motions) together with the radial velocity of any source have a very precise definition within a relativistic framework for astrometry \cite{K2003}. Several considerations like the gravitational light deflection by Solar System bodies or the fully relativistic form for the aberration formulas must be taken into account to achieve the desired accuracy. Morover, for obtaining the most precise modern ephemerides, provided by NASA's JPL Development Ephemerides, the Einstein-Infeld-Hoffmann equations are used in the so-called first post-Newtonian approximation of General Relativity to treat the $N$-body problem \cite{AGR}, which include corrections in inverse powers of the speed of light to the classical Newtonian equations. These highly accurate ephemerides will be used later in our simulations. 

Furthermore, it was proposed in $2016$ that there could be a ninth planet responsible for anomalous behaviours in some trans-Neptunian objects, and several investigations are currently under way (see \textit{e.g.} \cite{PN1, PN2, PN3, PN4} and references therein). Altough some constraints on its possible positions on the sky have been suggested, no telescopic search has proven successful to date, as it would indeed be an extremely faint object. However, several additional hypotheses for explaining these irregularities have been proposed as well, such as a primordial black hole in the outer Solar System \cite{PNBH} and the isotropy symmetries of extreme trans-Neptunian objects \cite{PNs}.

It is truly astonishing that all modern achievements and investigations, naturally accompanied by the exponential increase of technological developments, are being held only a century and a half later than the discovery of Neptune, who followed from perturbation techniques applied to basic Newtonian gravitational theory, by using just paper and pen. It is our goal in this paper to study this milestone in Astronomy from a different, numerical approach, outlined in Section \ref{Mathematical setup}. Our results are presented in Section \ref{UranusNeptune}. An additional analysis regarding the discovery of Pluto is also given in Section \ref{NeptunePluto}.

\section{Mathematical and computational setups}
\label{Mathematical setup}

\noindent
Until the advent of General Relativity in 1915, the dynamics of the gravitational $N$-body problem for point-like masses in Celestial Mechanics were classically given by combining Newton's second law of motion with his gravitational law, which nowadays we know is the general relativistic limit of non-relativistic velocities in the system and weak and slowly changing gravitational fields. Let $m_i$ be the mass of body $i$ with $i = 1, \dots, N$ and $\bold{r}_i$ its position vector with respect to the origin of an arbitrary inertial reference system. Assuming the equality between gravitational and inertial masses, \textit{i.e.} the equivalence principle, the dynamical equation for body $i$ reads
\begin{equation}
\frac{d^2 \bold{r}_i}{dt^2} = - G \displaystyle{\sum_{\substack{j = 1 \\ j \neq i}}^N} \frac{m_j}{||\bold{r}_i - \bold{r}_j||^3} (\bold{r}_i - \bold{r}_j) \hspace{0.1cm}; \quad i = 1, ..., N,
\label{NG}
\end{equation}
where $G$ is Newton's gravitational constant and $||\cdot||$ denotes the usual $3$-dimensional Euclidean norm. Equivalently, the dynamics of the system can be studied in the framework of Hamiltonian mechanics, whose structure is more suitable for implementing numerical schemes. In this formalism, equation (\ref{NG}) can be derived from the Hamilton equations corresponding to the Hamiltonian of the gravitational $N$-body system 
\begin{equation}
H_{N \text{ body}}\left(\{\bold{r}_i\}, \{\bold{p}_i\}\right) = \sum_{i = 1}^N \frac{||\bold{p}_i||^2}{2m_i} - G \sum_{i = 1}^{N - 1} \sum_{j = i + 1}^N \frac{m_i m_j}{||\bold{r}_i - \bold{r}_j||},
\label{genH}
\end{equation}
$\bold{p}_i$ being the corresponding conjugate momenta of body $i$. 

As is well-known, general systems with $N \geq 3$ are non-integrable and no exact analytical solution can be found, and even for the Keplerian problem with $N = 2$ it is not possible to obtain a solution as a function of time in closed form\footnote{Furthermore, chaotic behaviours appear for $N \geq 3$ in many situations after sufficiently large times. See for example Section $9.2$ of \cite{MD}.}. Consequently, one needs to resort to perturbative or numerical methods in order to make progress; the former approach was historically developed to study accurate planetary dynamics and, actually, many perturbation theory techniques in Classical Mechanics were developed due to their extensive necessity and use in Dynamical Astronomy \cite{goldstein}. In Celestial Mechanics of the Solar System, this approach considers the Keplerian motion of \textit{e.g. }a planet and the Sun and then takes into account the effect of further planets perturbatively, \textit{e.g. }by means of the corresponding disturbing function and its series expansions in terms of particular combinations of the orbital elements of the involved planets starting from an expansion in Legendre polynomials under some reasonable conditions \cite{MD}. It was the development of these techniques that, after remarkably long calculations, led to the prediction of the mass and position of Neptune at the end of the first half of the $19$th century, a truly memorable milestone in Science given that, at that time, not even basic calculators were available to perform numerical computations. 

The disturbing function appears naturally in the $3$-body problem after writing the equations of motion for the relative position vectors from the Sun to each planet. Considering (\ref{NG}) with $N = 3$ and defining $M_\star \equiv m_1$ (main, primary mass) and the relative position vectors $\bold{r}_{i\star} \equiv \bold{r}_i - \bold{r}_1$; $i = 2, 3$, one obtains
\begin{equation}
\frac{d^2 \bold{r}_{2\star}}{dt^2} + G ~ \frac{M_\star + m_2}{||\bold{r}_{2\star}||^3} ~ \bold{r}_{2\star} = \frac{\partial \mathcal{R}_{23}}{\partial \bold{r}_{2\star}} ~ ,  \quad \quad \quad \frac{d^2 \bold{r}_{3\star}}{dt^2} + G ~ \frac{M_\star + m_3}{||\bold{r}_{3\star}||^3} ~ \bold{r}_{3\star} = \frac{\partial \mathcal{R}_{32}}{\partial \bold{r}_{3\star}} ~ .
\label{3body equations}
\end{equation}
The left-hand sides include the terms corresponding to the Keplerian, $2$-body problems, whereas each right-hand side contains the information about the interaction with the third body, yielding a contribution to the Keplerian potential which is what we call the disturbing function:
\begin{equation}
\mathcal{R}_{23} \equiv G m_3 \left(\frac{1}{||\bold{r}_{3\star} - \bold{r}_{2\star}||} - \frac{\bold{r}_{3\star} \cdot \bold{r}_{2\star}}{||\bold{r}_{3\star}||^3}\right) ~ ,  \quad \quad \quad \mathcal{R}_{32} \equiv G m_2 \left(\frac{1}{||\bold{r}_{2\star} - \bold{r}_{3\star}||} - \frac{\bold{r}_{2\star} \cdot \bold{r}_{3\star}}{||\bold{r}_{2\star}||^3}\right) ~ .
\label{disturbing functions}
\end{equation}
Explicit expansions of the disturbing function in terms of the $6$ orbital elements of the involved planets as Le Verrier developed them can be found \textit{e.g.} in \cite{MurrayExp, MurrayNote}. Generally, in an $N$-body problem for $N \geq 3$, the interaction between each pair of planets can be written in terms of the disturbing function in the same way as shown in  \ref{3body equations} and \ref{disturbing functions}; for an arbitrary planet $i$, with $2 \leq i \leq N$, we have
\begin{equation}
\frac{d^2 \bold{r}_{i\star}}{dt^2} + G ~ \frac{M_\star + m_i}{||\bold{r}_{i\star}||^3} ~ \bold{r}_{i\star} = \sum_{\substack{j = 2 \\ j \neq i}}^N\frac{\partial \mathcal{R}_{ij}}{\partial \bold{r}_{i\star}} ~ ; \quad \quad \mathcal{R}_{ij} \equiv G m_j \left(\frac{1}{||\bold{r}_{j\star} - \bold{r}_{i\star}||} - \frac{\bold{r}_{j\star} \cdot \bold{r}_{i\star}}{||\bold{r}_{j\star}||^3}\right) ~ .
\label{full disturbing funct}
\end{equation}

On the other hand, with the advent of computers, a lot of work has been done during the last decades regarding the development of numerical methods to study many problems in all branches of Science and several techniques to treat the general $N$-body problem in this way can be implemented. This will be our approach in the present paper for studying the peturbations induced in the orbit of Uranus by Neptune. The system of equations (\ref{NG}) has then to be numerically integrated. For our results, we did so by means of the Wisdom-Holman $N$-body map, implemented in the Python package "REBOUND" \cite{rebound}, which is appropriate for systems in which there are not very close encounters between any of the bodies. The foundations for this symplectic integrator are next quickly outlined (see \cite{wh, reboundwhfast, MD} for a detailed discussion).

The general $N$-body Hamiltonian (\ref{genH}) needs first to be split into a sum of Keplerian Hamiltonians representing usual integrable individual $2$-body problems and a "small" interaction part describing the planet interactions in the $N$-body problem. This is achieved by introducing the so-called Jacobian coordinates, which are defined in what follows. Let $i = 1$ denote again the primary mass $M_\star \equiv m_1$, so that the remaining $N - 1$ bodies orbit around it. For each of these orbiting bodies one defines then its Jacobi coordinates as $\bold{r}_i^J \equiv \bold{r}_i - \bold{R}_{i - 1}$; $i = 2, \dots, N$, where $\bold{R}_i$ is defined as the position vector of the center of mass of the subsystem composed only by the first $i$ bodies, \textit{i.e.} $\bold{R}_i \equiv \frac{1}{M_i} \sum_{k = 1}^i m_k \bold{r}_k$, with $M_i \equiv \sum_{k = 1}^i m_k$, whereas the Jacobian coordinates for the primary mass are just those of the center of mass of the complete system, $\bold{r}_1^J \equiv \bold{R}_N$. The corresponding conjugate momenta read $\bold{p}^J_i = m^J_i \frac{d \bold{r}_i^J}{dt}$, where the masses $m^J_i$ are defined as $m^J_{i \geq 2} \equiv \frac{M_{i - 1}}{M_i} m_i$ and $m^J_1 \equiv \sum_{i = 1}^N m_i$ (total mass of the system); concretely, one has $\bold{p}^J_1 = \sum_{i = 1}^N \bold{p}_i$, \textit{i.e.} the momentum of the first Jacobian particle is just the momentum of the whole system. By means of these transformations and defining\footnote{This definition is formally introduced to make connection with the disturbing function formalism discussed earlier \cite{reboundwhfast}.} $\mathscr{M}^J_i \equiv \frac{M_i}{M_{i - 1}} m_1$; $i = 2, \dots, N$, it is shown that the Hamiltonian (\ref{genH}) can be brought into the form
\begin{equation}
H_{N \text{ body}} = H_\text{Free} + \sum_{i = 2}^N H_{\text{Keplerian}, ~ i} ~ + H_\text{Interaction} ~,
\label{split Hamiltonian}
\end{equation}
where 
\begin{equation}
H_\text{Free} \equiv \frac{||\bold{p}^J_1||^2}{2 m^J_1}, \quad H_{\text{Keplerian}, ~ i} \equiv \frac{||\bold{p}^J_i||^2}{2 m^J_i} - G ~ \frac{m^J_i \mathscr{M}^J_i}{||\bold{r}^J_i||}, \quad H_{\text{Interaction}} \equiv G \left(\sum_{i = 2}^N \frac{M_\star m_i}{||\bold{r}^J_i||} - \sum_{i = 1}^{N - 1} \sum_{k = i + 1}^N \frac{m_i m_k}{||\bold{r}_i - \bold{r}_k||}\right).
\label{hamiltonian terms}
\end{equation}
Now, $H_\text{Free}$ and each $H_{\text{Keplerian}, ~ i}$ are respectively the Hamiltonians of a free point-like mass (the $N$-body center of mass, as expected since we work in an inertial system) and an individual $2$-body system, in the Jacobian coordinates. On the other hand, the interaction Hamiltonian $H_{\text{Interaction}}$ is shown in (\ref{hamiltonian terms}) to be written in terms of both the usual and Jacobi coordinates; notice that it does depend neither on $\bold{p}_i$ nor on $\bold{p}^J_i$, so the coordinates are constants of motion under the application of $H_{\text{Interaction}}$ alone and, considered separately, each term in (\ref{split Hamiltonian}) is integrable.

The original Wisdom-Holman approach \cite{wh} consits of substituting the "true" interaction Hamiltonian in (\ref{split Hamiltonian}) by a mapping Hamiltonian\footnote{The delta distributions are to be understood in the sense that they will appear under an integral sign when solving Hamilton's dynamical equations.} 
\begin{equation}
H_\text{Mapping} \equiv 2 \pi H_{\text{Interaction}} \sum_{l \in \mathbb{Z}} \delta_\text{D}(\Omega_M t - 2 \pi l) = \tau H_{\text{Interaction}} \sum_{l \in \mathbb{Z}} \delta_\text{D}(t - \tau l),
\label{H mapping}
\end{equation}
where the frequency $\Omega_M$ is of the order of the planetary orbital frequencies and $\tau = \frac{2 \pi}{\Omega_M}$ is the corresponding time period. This substitution includes a periodic series of Dirac delta distributions denoted by $\delta_\text{D}(\cdot)$ which, for a small enough timestep $\tau$ (typically a small fraction of the smallest dynamical timescale in the problem under consideration), represent high frecuency terms that barely affect the long-term evolution of the system on average. For a detailed discussion of the induced error $H_\text{Mapping} - H_\text{Interaction}$ as a function of the timestep by this method see \textit{e.g.} \cite{SI}. The key idea of the previous replacement lies in that, between each pair of times in which the delta distributions do not vanish, the system evolves only under the integrable Hamiltonian terms $H_\text{Free} + \sum_{i = 2}^N H_{\text{Keplerian}, ~ i}$, whereas whenever there is a contribution from $\delta_\text{D}(t - \tau l)$, the system gets an impulse or "kick" under $H_\text{Mapping}$, whose corresponding Hamilton equations are also readily integrated since it does not depend on any of the momenta (a conversion between Jacobian and usual coordinates must be previously performed). This way, the evolution is controlled by successive applications of the Keplerian integrable problem and $\delta_\text{D}$-distribution "kicks". A detailed discussion on algebraic mappings can be found for instance in Chapter $9.5$ of \cite{MD}.

\section{Perturbations induced in the orbit of Uranus by Neptune}
\label{UranusNeptune}

\noindent
We now turn our attention to the problem of the irregularities the orbit of Uranus around the Sun (the primary mass, with $M_\star = M_\odot$) suffers as a result of its gravitational interaction with Neptune, following the numerical approach outlined in the previous section. In order to do this, we first consider the effects that Jupiter, Saturn and Neptune separately have on the $2$-body problem of Uranus and the Sun and then we study how the motion of Uranus is perturbed when all the planets except Neptune are included. To initialize the simulations, we must give initial conditions for each body; these, as well as the necessary masses, are taken from the JPL Horizons system \cite{JPL}, at some given initial times\footnote{These ephemerides are given in Baricentric Dynamical Time, $t = TDB$. For our computations this is not relevant, but it has to be taken into account \textit{e.g.} when doing relativistic astrometry.} we specify later in each case. The ecliptic plane is considered at the J$2000.0$ standard epoch.

For all our simulations in this paper we chose a timestep of $\tau = \frac{10^{-4}}{2 \pi} \text{ yr}$, which is much smaller than the smallest orbital period in every case, so that numerical errors are negligible as no improvement is achieved when decreasing more the timestep. The Python script returns in each case the ecliptic Cartesian coordinates of each body $i$ from the Solar System barycenter, $\bold{r}^\text{B, ecl}_i$, and the corresponding velocities, at the integration times we specify. With these, we can readily evaluate the geocentric equatorial Cartesian coordinates $\bold{r}^\text{G, eq}_i$ of body $i$ by means of the classical equation
\begin{equation}
\bold{r}^\text{G, eq}_i = \mathcal{R}(\epsilon) \cdot \left(\bold{r}^\text{B, ecl}_i - \bold{r}^\text{B, ecl}_\text{Earth}\right); \quad \quad \mathcal{R}(\epsilon) \equiv \begin{pmatrix}
1 & 0 & 0\\
0 & \cos \epsilon & - \sin \epsilon\\
0 & \sin \epsilon & \cos \epsilon
\end{pmatrix},
\label{Coordinate transformation}
\end{equation}
where $\mathcal{R}$ represents the rotation from the ecliptic to the equatorial plane of an angle $\epsilon$ (Earth's axial tilt, currently $\epsilon \simeq 23.44^\circ$) around the $x$-axis, which points towards the vernal equinox in both cases. From these, we readily evaluate the geocentric right ascensions and declinations correspondingly.

\begin{figure}[H]
	\centering
	\hspace*{-1cm}\includegraphics[width = 1.15\linewidth]{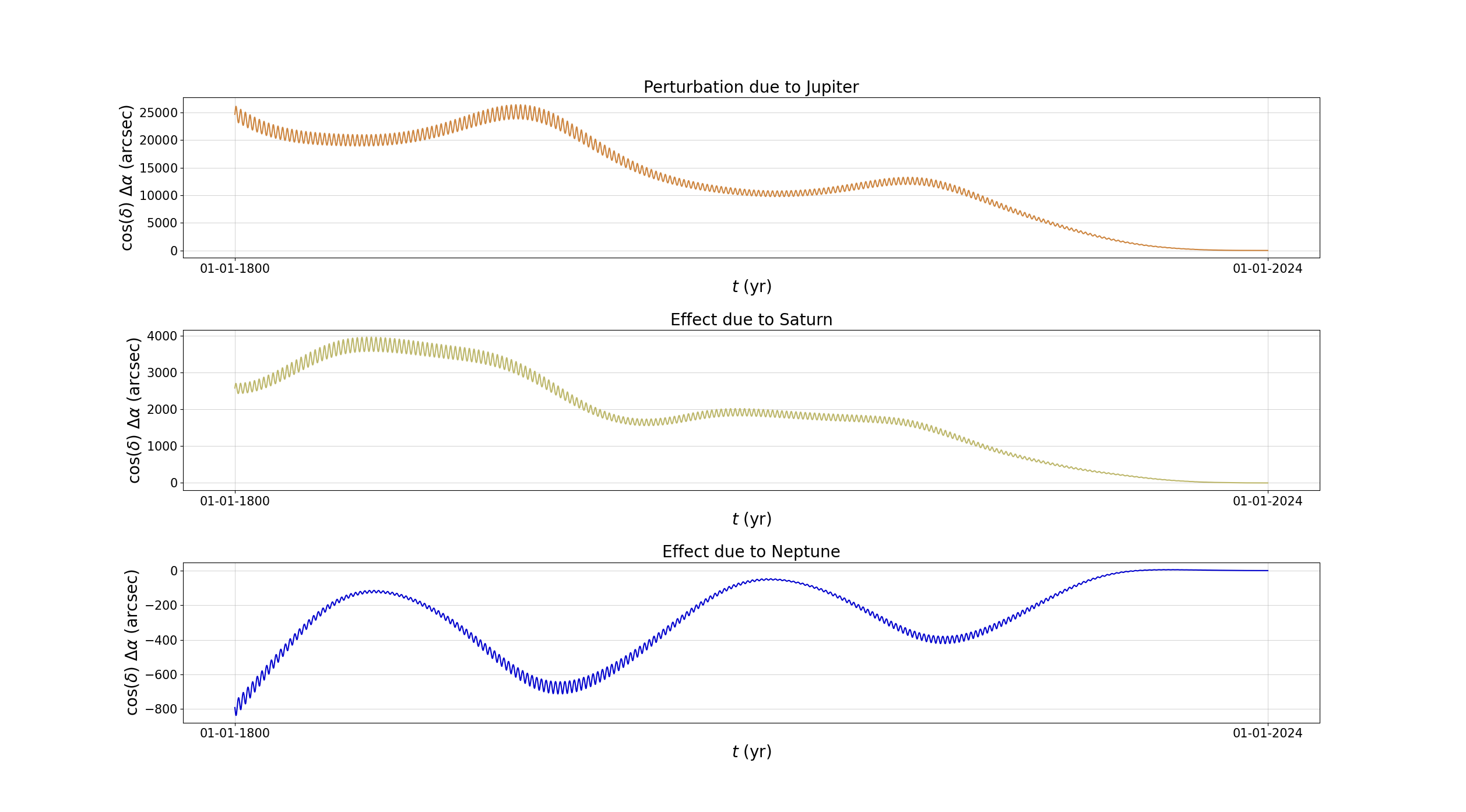}
	\caption{Shifts in the right ascension of Uranus due to perturbations of the other giant planets, starting the simulations at $TDB = 01$-$01$-$2024$ $00$:$00$ and integrating backwards in time.}
	\label{UranusRAS}
\end{figure}

\begin{figure}[H]
	\centering
	\hspace*{-1cm}\includegraphics[width = 1.15\linewidth]{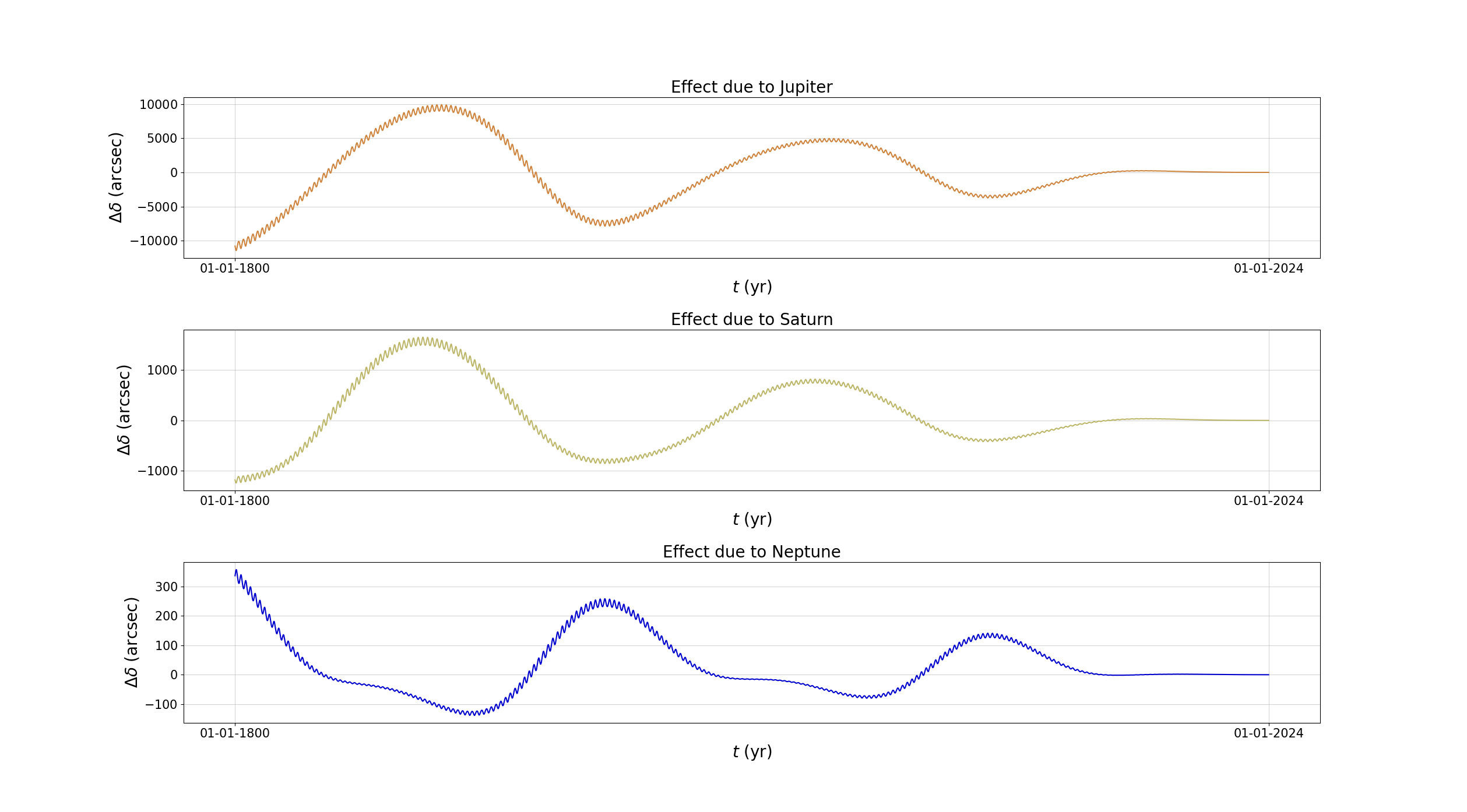}
	\caption{Shifts in the declination of Uranus due to perturbations of the other giant planets.}
	\label{UranusDECS}
\end{figure}

In Figures \ref{UranusRAS} and \ref{UranusDECS} we show respectively the contribution of each giant planet separately to the geocentric right ascension and declination of Uranus as it follows a Keplerian motion around the Sun. For this purpose, we initialize in each case two simulations at the $TDB$ date $t_0 = 01$-$01$-$2024$ $00$:$00$, one of which implements Uranus and the Sun with the corresponding JPL ephemerides and another one which also includes the corresponding giant planet, and integrate backwards until some past date, in our case $01$-$01$-$1800$ $00$:$00$. Equation (\ref{Coordinate transformation}) is then used to obtain the equatorial coordinates of Uranus in both cases, whose difference yields the right ascension and declination shifts resulting from the perturbation of each giant planet. Denoting by $(\alpha, \delta)$ and $(\Tilde{\alpha}, \Tilde{\delta})$ the equatorial coordinates with and without the perturber respectively, we write the shifts as
\begin{equation}
\Delta \alpha (t) \equiv \Tilde{\alpha}(t) - \alpha(t) ~ , \quad \quad \Delta \delta (t) \equiv \Tilde{\delta}(t) - \delta(t) ~ .
\label{Shifts}
\end{equation}
Since both simulations of each pair are started at the same date, we always have $\Delta \alpha (t_0) = \Delta \delta (t_0) = 0$.

In each plot we note overall periodic oscillations due to the relative position between Uranus and each corresponding giant planet which depend on their orbital parameters, as well as annual oscillations due to the motion of the geocenter around the Sun\footnote{The geocenter is not included in these simulations to avoid adding further perturbations of any kind to the corresponding $2$ and $3$-body problems; instead, its position as a function of the simulation time, $\bold{r}^\text{B, ecl}_\text{Earth} (t)$, is obtained via a sepparate simulation with the same parameters which includes all the known Solar System planets, so that it returns much more accurate positions.}, as happens with the parallactic ellipse of distant sources. As expected, the perturbations by Jupiter are the most noticeable ones, followed by those due to Saturn and finally by Neptune. As time goes by, both configurations in each pair of simulations diverge from each other, resulting in an increasing amplitude of the overall oscillations.

\begin{figure}[H]
	\centering
	\hspace*{-1cm}\includegraphics[width = 1.10\linewidth]{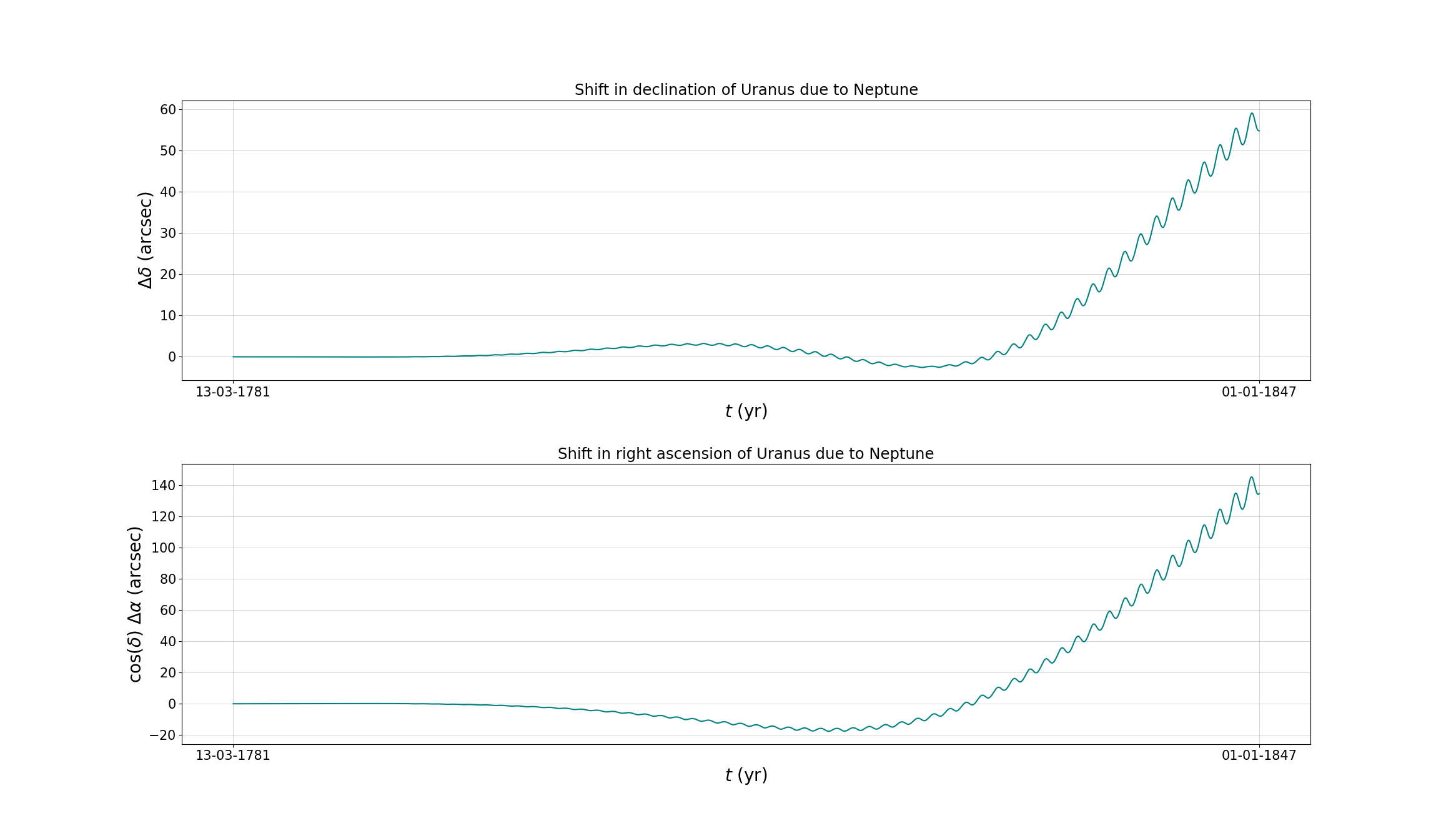}
	\caption{Variations in the geocentric equatorial coordinates of Uranus due to Neptune, starting the simulations in the day $13$-$03$-$1781$, when Uranus was officially discovered, until $01$-$01$-$1847$, after Neptune was discovered.}
	\label{UrG}
\end{figure}

\begin{figure}[H]
	\centering
	\hspace*{-1cm}\includegraphics[width = 1.10\linewidth]{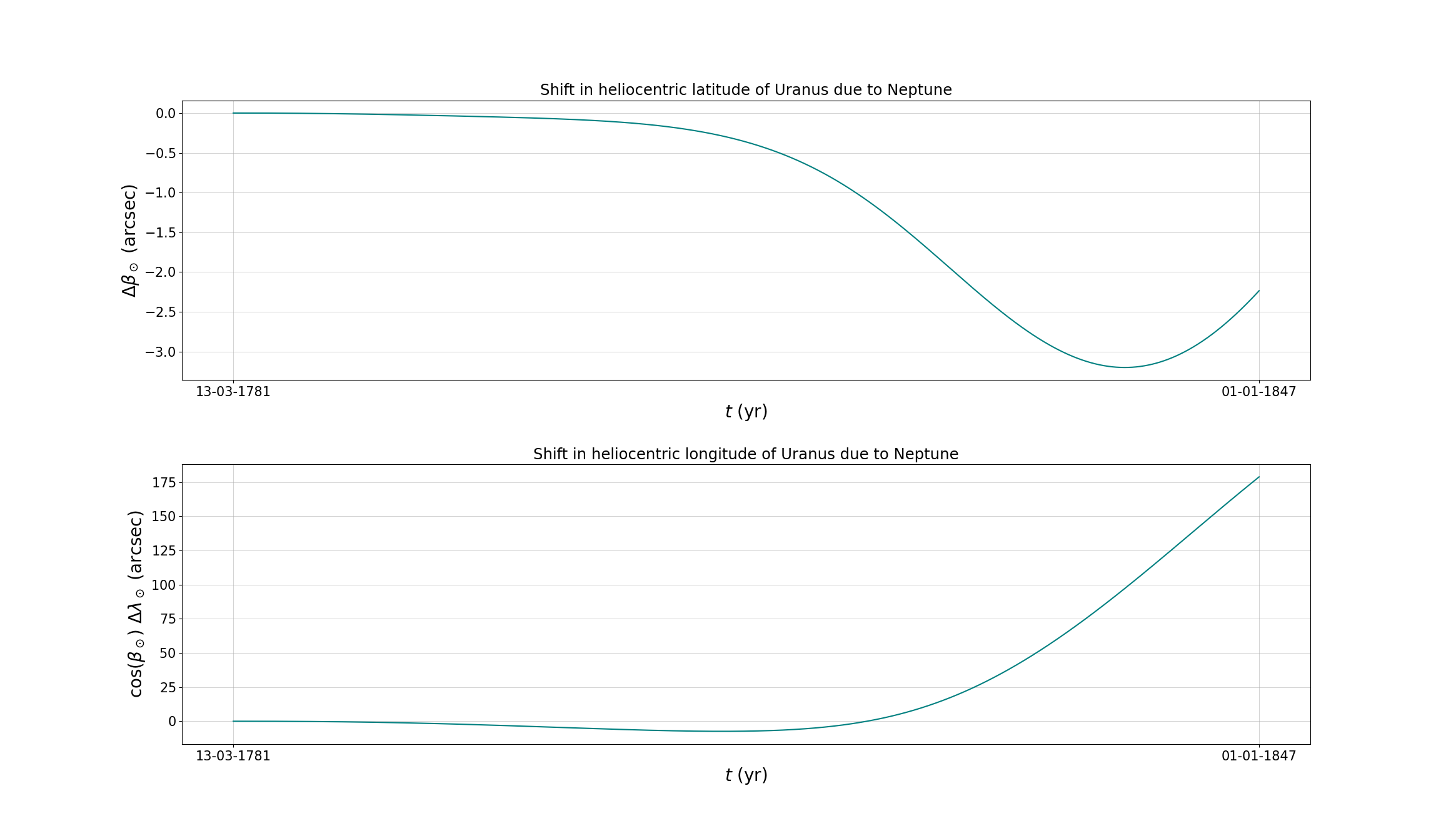}
	\caption{Perturbations in the heliocentric ecliptic coordinates of Uranus due to Neptune, again from $13$-$03$-$1781$ to $01$-$01$-$1847$.}
	\label{UrCOM}
\end{figure}

Historically, only post-discovery observations were used by Bode to build the first tables of Uranus in $1821$ and to try to predict its orbit, as pre-discovery ones (found to be no less than $19$, the first recorded observation made by the first Astronomer Royal John Flamsteed in $1690$, who mistook it for a star, "$34$ Tauri") were thought to have too much observational uncertainty \cite{Smart}. For this reason, in order to determine the irregularities on the orbit of Uranus by Neptune from the day of its official discovery, on March $13$, 1781, we initialize again two simulations at time $t_0 = 13$-$03$-$1781$ $00$:$00$, both containing all the planets of the Solar System from Mercury to Uranus (we also include the inner planets for completeness) and one of which also includes Neptune, and let them evolve until the late $1840$s. Besides computing the shifts in geocentric right ascension and declination of Uranus as in the previous cases, we also evaluate for clarity the shifts in heliocentric (more precisely, barycentric, although these small differences do not modify our results) ecliptic coordinates  
\begin{equation}
\Delta \lambda_\odot (t) \equiv \Tilde{\lambda}_\odot(t) - \lambda_\odot(t) ~ , \quad \quad \Delta \beta_\odot (t) \equiv \Tilde{\beta}_\odot(t) - \beta_\odot(t) ~ ,
\label{Shifts helio}
\end{equation}
where in this case $(\Tilde{\lambda}_\odot, \Tilde{\beta}_\odot)$ are the heliocentric ecliptic longitude and latitude of Uranus when Neptune is absent from the system, and $(\lambda_\odot, \beta_\odot)$ are the analogs when the trans-Uranian planet is included. The corresponding results are shown in Figures \ref{UrG} and \ref{UrCOM}, respectively. Of course, the annual oscillations are not present in the latter as no orbital motion of Earth is involved in these coordinates.

In both Figures, we notice remarkable deviations which were indeed observable at the beginning of the $19$th century. By $1832$, the observed mismatch in heliocentric longitude was $|\Delta \lambda_\odot (TDB = 1832)| \simeq \frac{1}{2}'$, and Airy pointed out in $1837$ that these errors were dramatically increasing \cite{Smart}, behaviours which agree with our numerical results in the lower plot of Figure \ref{UrCOM}. Notice that the deviations in ecliptic latitude (upper plot) are much smaller, since in both configurations the motion of Uranus takes place nearly on the same plane, as all the Solar System planets orbit with small inclinations with respect to the ecliptic plane.

\section{Effects in the orbit of Neptune due to Pluto}
\label{NeptunePluto}

\noindent
After the discovery of Neptune, Percival Lowell, among others, speculated that a ninth planet could be further perturbing the orbits of Neptune and Uranus, so he organized and started exhaustive investigations and searches at the Lowell Observatory at the beginning of the $20$th century. The current dwarf planet Pluto would be discovered at this place later in 1930 by Clyde Tombaugh. Nowadays, it is known that this discovery occurred by chance, since the effects it induces in the orbits of the ice giants are so small that these could not represent the deviations claimed \textit{e.g.} by Lowell. Following the same procedure as in Section \ref{UranusNeptune}, we now analyze the effect that the Pluto-Charon system has on Neptune to show that this is indeed the case. The perturbations on Uranus are consistently much smaller.

\begin{figure}[H]
	\centering
	\hspace*{-0.65cm}\includegraphics[width = 1.1\linewidth]{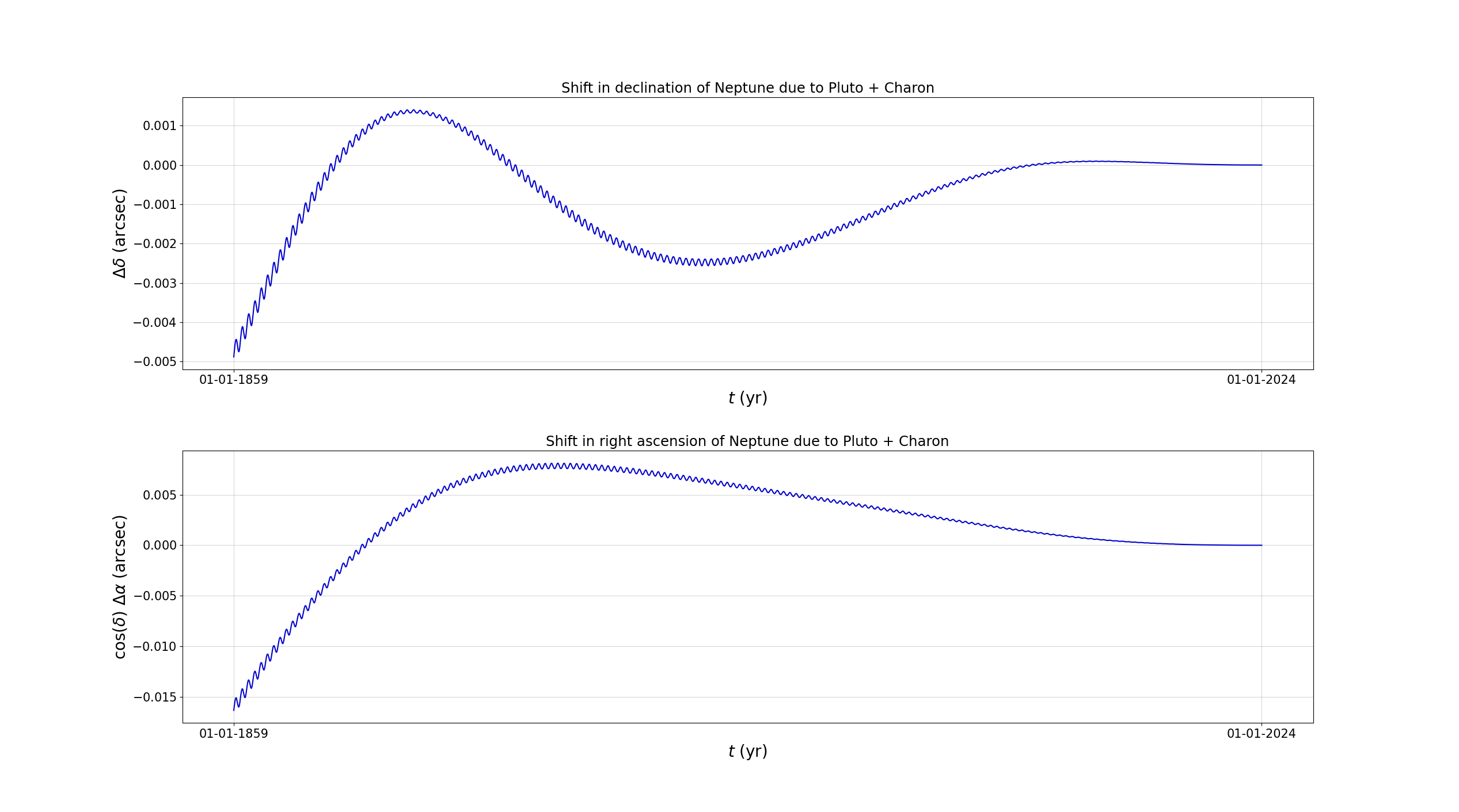}
	\caption{Shifts in the geocentric right ascensions and declinations of Neptune if Pluto was absent, starting the simulations in $01$-$01$-$2024$ and integrating backwards during a complete Neptune orbital period.}
	\label{Neptune and Pluto}
\end{figure}

In this case, the simulations shown in Figure \ref{Neptune and Pluto} are started again at the beginning of the present year,  $t_0 = 01$-$01$-$2024$ $00$:$00$ and integrated backwards to see the accumulated variations in Neptune's right ascension and declination over one orbital period of Neptune, \textit{i.e.} until $1859$. All known planets of the Solar System are included in both runs.

We notice that the corresponding shifts in the equatorial coordinates of the eighth planet only reach a few milliarcseconds in the time span of one Neptune orbital period. These deviations may in fact be smaller than those coming from some relativistic effects which must be taking into account \textit{e.g.} in high-precision astrometry, but which were unaccounted for at the very beginning of the $20$th century. Hence, this approach rules out the possibility that the Pluto-Charon system was responsible for the alleged irregularities in the orbits of the ice giants.

As a matter of fact, the mass $m_P$ that a trans-Neptunian object should have to induce in the orbit of Neptune the same observable effects that the latter body induced in the orbit of Uranus after a time span of $165$ years can be estimated along the same lines as our previous approaches. With the same parameters as the previous simulation and assuming the distance of this supossed perturber to be given by the next-to-Neptune power in the Titius-Bode law for this estimation, $||\bold{r}_p|| \sim 0.4 + 0.3 \cdot 2^8$ AU $= 77.2$ AU, and that it follows a circular orbit, it must be verified that
$$
m_P \gtrsim 4\cdot 10^{-5} M_\odot \sim 0.04 ~ M_\text{Jupiter}.
$$
With this comment we conclude our modern approach to the problem of the historical Solar System planetary perturbations.

\section{Final remarks}

By using a modern, numerical approach of the gravitational $N$-body problem by means of the Wisdom-Holman map, we were able to successfully study the problem of the irregularities in the orbit of Uranus due to Neptune, revisiting the milestone made by Adams and Le Verrier. The effect of each giant planet on the orbit was also separately analyzed; as was to be expected, the shifts in the position of Uranus due to Jupiter and Saturn were much greater than that caused by Neptune, exceeding even $1^\circ$ in the geocentric equatorial coordinates after a relatively short period of time. Finally, we considered the perturbation that the Pluto-Charon system inflicts on Neptune, sustaining that the reported discrepances in the orbits of the ice giants at the beginning of the last century could not be produced by it. The method we followed can be employed in general to study the deviation in the position of any body due to a external perturber during a given time interval.

Nowadays, we know that the Solar System is much more intrinsically complex than astronomers perceived it to be until the $20$th and $21$th centuries. Accordingly, the description of many subtle effects in observational data from modern missions requires highly sophisticated models, some of which involve, for instance, general relativistic corrections, as was pointed out in the text, as well as highly refined statistical methods. Interestingly, it remains one of the most intriguing modern unknowns in Astronomy to explain the alleged irregularities in some trans-Neptunian objects which, as pointed out in Section \ref{Intro}, suggest among other hypotheses the existence of a $9$th planet in our Solar System, more than $150$ years after the discovery of Neptune.

\subsection*{Acknowledgements}

\noindent
The authors would like to thank Davor Krajnović from the Leibniz-Institut für Astrophysik Potsdam, the staff at the Académie des sciences and Jorge Núñez from the Observatori Fabra for helpful correspondence during the preparation of this work, as well as Antonio Elipe from the University of Zaragoza for carefully reading the mansucript. GRM is also grateful to Enrico Gerlach from the Lohrmann Observatory for pointing out the availability of the "REBOUND" Python package. This project has received funding from the European Union’s Horizon 2020 research and innovation programme under the Marie Skłodowska-Curie grant agreement No 101072454 (MWGaiaDN). It was also supported by the Spanish "Ministerio de Ciencia y Innovación" under the Project PID 2021-122608NB-IOO (AEI/FEDER/UE).

\printbibliography

\end{document}